\documentclass[letterpaper,aps,prd,reprint,showpacs,floatfix,amsmath,amssymb]{revtex4-1}
\usepackage{graphicx}
\usepackage{color}
\usepackage{dcolumn}\usepackage{bm}
\usepackage{verbatim}

\newcommand{\ud}{\,\mathrm{d}}
\newcommand{\ue}{\,\mathrm{e}}

\newcommand{\h}{\cosh2\sqrt{3}\beta_-}

\newcommand{\hpm}{\ue^{-T}\cosh{\sqrt{3}\beta_{-0}}\pm\ue^{T}\sinh{\sqrt{3}\beta_{-0}}}

\begin{document}

\title{Mixmaster Revisited: Wormhole Solutions to the Bianchi IX Wheeler-DeWitt Equation using the Euclidean-Signature Semi-Classical Method}
\author{Joseph H. Bae}
\affiliation{Department of Physics, Yale University, New Haven, CT 06520, USA.}
\date{\today}
\begin{abstract}
A modified semi-classical method is used to construct both ground and excited state solutions to the canonically quantized vacuum Bianchi IX (Mixmaster) cosmological models. Employing a modified form of the semi-classical Ansatz we solve the relevant Wheeler-DeWitt equation asymptotically by integrating a set of linear transport equations along the flow of a suitably chosen solution to the corresponding Euclidean-signature Hamilton-Jacobi equation. For the Moncrief-Ryan (or `wormhole') Hamilton-Jacobi solution, we compute the ground state quantum correction term associated with operator ordering ambiguities and show how higher order correction terms can be computed. We also determine the explicit, leading order forms of a family of excited states and show how to compute their quantum corrections as smooth, globally defined functions on the Bianchi IX minisuperspace. These excited state solutions are peaked away from the minisuperspace origin and are labeled by a pair of positive integers that can be plausibly interpreted as graviton excitation numbers for the two independent anisotropy degrees of freedom. The Euclidean-signature semi-classical method used here is applicable to more general models, representing a significant progress in the Wheeler-DeWitt approach to quantum gravity.
\end{abstract} 
\pacs{04.60.-m,04.60.Ds,04.60.Kz}
\maketitle

\section*{Introduction}
The vacuum diagonalized Bianchi IX cosmological models (or the Mixmaster models) were first studied by Belinsky, Khalatnikov and Lifshitz (BKL) to investigate cosmological singularities~\cite{BKL}; and also by Misner, who used the Hamiltonian approach for the study of its dynamics and was the first to consider its quantization~\cite{Misner,Misner72}. Classically, the Mixmaster models are characterized by a point in $(\alpha, \beta_+, \beta_-)$-space, where $\{\alpha, \beta_+, \beta_-\}$ are the Misner variables. The scale parameter $\alpha(t)$ gives the size of the three-dimensional hyper surface relative to its initial size $l_0$; the anisotropy parameters $\beta_+(t)$ and $\beta_-(t)$ describe the anisotropy of the hyper surface. Together, they define the Bianchi IX minisuperspace. The classical trajectory of the system in the minisuperspace is  fraught with caustics and is thought to be chaotic \cite{Berger,Motter}. This presents a serious problem in trying to solve the relevant Wheeler-DeWitt equation by conventional semi-classical methods. Moreover, there is an operator ordering ambiguity once the canonical variables and momenta are promoted to quantum operators. Bianchi IX models have been studied in the contexts of homogeneous relativistic cosmologies and supergravity \cite{RyanShepley,Graham,Graham93,Graham93b,BeneGraham94,Graham94,Moniz1,DEath,Csordas95,Damour14}, in Euclidean signature settings \cite{GibbonsPage}, or with local approximations to the full Bianchi IX  potential \cite{Marolf}. So far, there has not been a semi-classical quantization scheme for the vacuum Bianchi IX models that both accommodates operator ordering ambiguities and yields globally defined functions as fundamental solutions to the Hamilton-Jacobi equation and its quantum corrections. In this paper, we employ a modified semi-classical method to calculate ground and excited state solutions and their quantum correction terms that are both globally defined on the Bianchi IX minisuperspace and are valid for a class of operator ordering ambiguities in the Wheeler-DeWitt equation. 

This paper is organized as follows: first, we review the canonical quantization of the diagonalized Bianchi IX models in the absence of matter fields, outlining how we obtain the Hamiltonian constraint and the relevant Wheeler-DeWitt equation. Next, we present the modified semi-classical method as developed by Moncrief et al. \cite{MMM}, and show how it can be modified further to apply to the vacuum Bianchi IX Wheeler-DeWitt equation. Choosing the Moncrief-Ryan (or `wormhole') solution of the Euclidean-signature Hamilton-Jacobi equation, we outline the computation of quantum correction terms to the ground and excited state solutions. The more mathematically technical properties of these quantum corrections---their existence to all orders in $\hbar$ and global smoothness on the associated minisuperspace---are under independent investigation, through the use of microlocal analytical methods \cite{Dimassi99}, by V. Moncrief \cite{VM14}.

\section{Method}
\subsection{Diagonalized Bianchi IX Line Element}
We set $c=1$ but retain all powers of $\hbar$ and $G$ throughout. In these units, the Planck length satisfies $l_P{}^2 = \hbar G$. It is well-known that in the absence of matter sources for the Einstein equations, the Mixmaster spacetime metric can always be put into diagonal form. Write the diagonalized vacuum Bianchi IX line element using Misner variables $\{\alpha, \beta_+, \beta_-\}$ \cite{Misner}: 
\begin{equation}
\begin{split}
\ud s^2 =& {}^{(4)}\!g_{\mu \nu} \ud x^\mu \ud x^\nu \\
\ud s^2 =& -N^2 \ud t^2 + \frac{l_0{}^2}{6\pi} \ue^{2\alpha}\big(\ue^{2\beta}\big)_{ij} \sigma^i \sigma^j,
\end{split}
\end{equation}
where $\{x^\mu\}=\{t,\theta,\phi,\psi\}$ with $t \in \mathbb{R}$; $N(t)$ is the lapse function; $\big(\ue^{2\beta}\big)_{ij}$ is a $3\times3$ matrix $\mathrm{diag}(\ue^{2\beta_+ + 2\sqrt{3}\beta_-},\ue^{2\beta_+ - 2\sqrt{3}\beta_-},\ue^{-4\beta_+ })$; and $\sigma^i$'s are the invariant differential one-forms on the $\mathbb{S}^3$ manifold, satisfying $\ud \sigma^i = \frac{1}{2}\epsilon_{ijk}\sigma^j \wedge \sigma^k$ \cite{Ellis}. The one-forms can be represented using 3-1-3 Euler angles $\{ \theta,\phi,\psi \}$ \cite{RyanShepley}: 
\begin{equation}
\begin{split}
\sigma^1 =&  \cos{\psi} \ud \theta + \sin{\psi}\sin{\theta}\ud \phi \\
\sigma^2 =&  \sin{\psi} \ud \theta - \cos{\psi}\sin{\theta}\ud \phi \\
\sigma^3 =&  \ud \psi + \cos{\theta}\ud \phi.
\end{split}
\end{equation}
From the line element above, we compute the canonical 4-volume measure and the scalar curvature of the metric ${}^{(4)}\!g_{\mu \nu}$:
\begin{eqnarray}
\sqrt{-g} =N \sin{\theta}  \, r^3 \! \ue^{3\alpha},\mspace{192mu} \\
\begin{split}
{}^{(4)}\!R  =& -\frac{6 \dot{N} \dot{\alpha}}{N^3}+\frac{6}{N^2}\Big( 2\dot{\alpha}^2+\ddot{\alpha} +\dot{\beta}_{+}{}^2+\dot{\beta}_{-}{}^2\Big)\\&-\frac{\ue^{-2\alpha}}{2r^2}\bigg( \mathrm{Tr}\Big(\ue^{4\beta}\Big) - 2 \mathrm{Tr}\Big(\ue^{-2\beta}\Big)\bigg),
\end{split}
\end{eqnarray}
where for convenience, we have defined $r = \frac{l_0}{\sqrt{6\pi}}$. We will see later that the last term of ${}^{(4)}\!R$ gives rise to the `potential term' $V(\beta_\pm)$ in the Wheeler-DeWitt equation:
\begin{equation}
\begin{split}
V(\beta_\pm)=&  \tfrac{1}{3}\Big( \mathrm{Tr}\big(\ue^{4\beta}\big) - 2 \mathrm{Tr}\big(\ue^{-2\beta}\big)\Big) \\
=&\tfrac{1}{3}e^{-8\beta_+}-\tfrac{4}{3}\ue^{-2\beta_+}\cosh{2\sqrt{3}\beta_-}\\&+\tfrac{2}{3}\ue^{4\beta_+}\left(\cosh4\sqrt{3}\beta_--1\right).
\end{split}
\end{equation}
We note  here that in some other works (for example, in \cite{RyanShepley}) the definition of $V(\beta_\pm)$ differs by $+1$.

\subsection{ADM Hamiltonian Formulation}
Write the Einstein-Hilbert action, evaluated on domains of the form $\Omega=\mathbb{S}^3\times {I}$, with ${I}:= [a,b] \subset \mathbb{R}$:
\begin{equation}
S_{EH} \,\,\,=\,\,\, \frac{1}{16\pi G}\int_{\Omega} \!\!\!\ud^4x \sqrt{-g}\,\,\,{}^{(4)}\!R. 
 %\,\,\,= \,\,\, \frac{\pi r^3}{G} \int \!\!\! \ud t\,\, N \ue^{3\alpha} \,\,\, {}^{(4)}\!R.
\end{equation}
Since the integrand is a function of $t$ only, we can carry out the integration over the angular coordinates $\{\theta,\phi,\psi\}$:
\begin{eqnarray}
\begin{split}
\int_{\Omega} \!\!\!\ud^4x \sqrt{-g}  {}^{(4)}\!R  = & \,\, r^3 \int_{\mathbb{S}^3} \sin{\theta} \ud \theta \ud \phi \ud \psi \int_{I} N \ue^{3\alpha} \, {}^{(4)}\!R \ud t \\ = & \,\, 16 \pi^2 r^3  \int_I N \ue^{3\alpha} \,\, {}^{(4)}\!R \ud t .
%\int \sin{\theta} \ud \Omega & = \int^\pi_0 \sin{\theta}\ud\theta \int^{2\pi}_{0}\ud \phi \int^{4\pi}_{0}\ud \psi = 16 \pi^2 .
\end{split}
\end{eqnarray}
Integrating by parts the $\ddot{\alpha}$ term in ${}^{(4)}\!R$, we write the ADM action that differs from the Einstein-Hilbert action by an additive boundary term \cite{ADM,MTW}:
\begin{eqnarray}
\begin{split}
S_{EH} =& S_{ADM} + \bigg[ \frac{\dot{\alpha} \ue^{3\alpha}}{N}\bigg]^b_a \\
S_{ADM}  =& \frac{\pi r^3}{G} \int_I \!\!\! \ud t\,\,  \bigg\{  \frac{6\ue^{3\alpha}}{N} \Big( -\dot{\alpha}^2 + \dot{\beta_+}^2 + \dot{\beta_-}^2 \Big) \\&- \frac{N \ue^{\alpha}}{2\,r^2}\bigg( \mathrm{Tr}\Big(\ue^{4\beta}\Big) - 2 \mathrm{Tr}\Big(\ue^{-2\beta}\Big)\bigg)  \bigg\}.
\end{split}
\end{eqnarray}
Constructing the Lagrangian from $S_{ADM}=\int L \ud t $, we define conjugate momenta $p_{\alpha}$ and $p_\pm$ (dimension $[\hbar]$) in the usual way:
\begin{equation}
\begin{split}
p_{\alpha} & = \frac{\partial L }{\partial \dot{\alpha}} = - \frac{12\pi \,r^3\! \ue^{3\alpha}}{NG} \dot{\alpha} \\
p_{\pm} & = \frac{\partial L }{\partial \dot{\beta_\pm}} =  \frac{12\pi \,r^3\! \ue^{3\alpha}}{NG} \dot{\beta}_\pm .
\end{split}
\end{equation}
Varying the ADM action (expressed in terms of the Hamiltonian $H_\perp$) with respect to the lapse $N$ gives us the Hamiltonian constraint equation $H_\perp = 0$:
\begin{eqnarray}
S_{ADM} = & \int  \ud t \,\, \big\{  p_\alpha \dot{\alpha} + p_+ \dot{\beta}_+ + p_- \dot{\beta}_- - N H_{\perp}  \big\} , \\
H_{\perp} = & \frac{G \ue^{-3\alpha}}{24 \pi \, r^3} \Big[\mathbf{p}\cdot\mathbf{p} + \frac{l_0{}^4}{G^2} \ue^{4\alpha} V(\beta_\pm) \Big] = 0. 
\end{eqnarray}
Note that we use the $(-++)$ `super-metric' notation to simplify notation throughout. In this notation, the wave operator is:
\begin{equation}
\Box = - \frac{\partial^2}{\partial \alpha^2} + \frac{\partial^2}{\partial \beta_+^2} +  \frac{\partial^2}{\partial \beta_-^2}
\end{equation}

\subsection{The Wheeler-DeWitt Equation}

Following Dirac \cite{Dirac58, Dirac82}, we promote the Hamiltonian constraint to an operator acting on wave function $\Psi$, and write the Wheeler-DeWitt Equation for the diagonalized Bianchi IX models:
\begin{equation}
\hat{H}_{\perp}\Psi= \frac{G\ue^{-3\alpha}}{24\pi r^3}\Big[\hat{\mathbf{p}}\cdot\hat{\mathbf{p}} +  \frac{l_0{}^4}{G^2} \ue^{4\alpha}V(\beta_{\pm})\Big]\Psi= 0.\label{WdW}
\end{equation}
As $\hat{p}_{\alpha},\hat{p}_{+}$, and $\hat{p}_{-}$ are operators realized by $-i\hbar\partial_\alpha$, $-i\hbar\partial_{\beta_+}$, and $-i\hbar\partial_{\beta_-}$ respectively, the factor $\ue^{-3\alpha}$ may be ordered with $\hat{p}_{\alpha}$ in different ways. We follow the ordering parametrization first suggested by Hartle and Hawking \cite{HHF}, and write $\ue^{-3\alpha}\hat{p}^2_{\alpha}$ as:
\begin{equation}
\begin{split}
\ue^{-3\alpha}\hat{p}^2_{\alpha}&=-\hbar^2\ue^{-(3-B)\alpha}\partial_{\alpha}\Big(\ue^{-B\alpha}\partial_\alpha\Big)\\&=-\hbar^2\ue^{-3\alpha}\partial^2_\alpha+\hbar^2B\ue^{-3\alpha}\partial_\alpha.
\end{split}
\end{equation}
Leaving in the operator ordering parameter $B$ means we can study a family of different operator orderings at once; later, we will see that the quantum correction terms to the ground and the excited state solutions depend critically on the value of $B$. 

Before we delve into presenting our Modified Semi-classical quantization scheme, we wish to set our current work in its context. The canonical formalism was set up in the late 1960s by the seminal works of B. DeWitt, C. Misner, and J. Wheeler \cite{ADM,DeWitt67,Wheeler70}. In writing the Wheeler-DeWitt equation, they constructed the wave function of the model universe $\Psi$ and the constraint equation that $\Psi$ must satisfy due to the underlying coordinate invariance of gravity (we will not talk about the three momentum constraints in this paper, as we are dealing with diagonalized vacuum Bianchi IX models). The Wheeler-DeWitt equation (the quantized Hamiltonian constraint) is a second-order partial differential equation for the wave function that expresses time-reparametrization invariance. Its form seems to imply that the wave function is static; this is the `problem of time' in canonical quantum cosmology. One of the ways of working around this problem is to conclude that the actual temporal evolution must be measured, not with respect to some external `time' parameter, but rather with respect to some internal `clock' variable contained within the system. In other words, the Wheeler-DeWitt equation is to be interpreted in terms of correlations between a `system' and a `clock'. Much work has been done in semi-classical cosmology by applying this approach to $k=1$ FRW models coupled to a massive scalar field \cite{Banks85,Kiefer87,Brout89,Bertoni96,Kim97}. In the case of our vacuum Bianchi IX universes, the most obvious candidate for such a clock variable is the dimensionless scale parameter $\alpha$. Classically, $\alpha$ evolves in a monotonic fashion from the `big bang' at a finite time in the past (when $\alpha =-\infty$), to a moment of maximal volume (when $\dot{\alpha} =0$), and again decreases monotonically to a `big crunch' at a finite time in the future (when again $\alpha =-\infty$) \cite{LinWald,Rendall97}. Thus, $\alpha$ could act as a clock variable as long as we can limit our analysis to the expansion phase.

However, the Mixmaster Wheeler DeWitt equation does not have Schr\"{o}dinger form; rather, it takes the form of a Klein-Gordon type equation, in that it is second order in all variables including $\alpha$, our clock variable, with a Lorentzian `super-metric'. This means that many of the usual constructions of quantum mechanics such as eigenvalues and eigenfunctions, and the conservation in `time' of `probability measures' do not apply. When all of our `eigenvalues' of Eq.~(\ref{WdW}) are identically zero, can we expect to find the discrete solutions of the Mixmaster Wheeler-DeWitt equation? This was a question asked some fifty years ago; and one of the main results of this paper is to show the spectrum of discrete solutions explicitly by applying the modified semi-classical approach to the vacuum Bianchi IX Wheeler-DeWitt equation.

\subsection{Modified Semi-Classical Method}
In their 2012 paper \cite{MMM} on ``Modified Semi-Classical Methods for Nonlinear Quantum Oscillations Problems'', Moncrief et al. develop a modified semi-classical approach to find the approximate solutions of the Schr\"{o}dinger equation for systems with nonlinear quantum oscillators. This method involves writing the Ansatz for the ground state wave function in the form:
\begin{equation}
\Psi = \ue^{-S_{\hbar}/\hbar}, \label{Ansatz1}
\end{equation}
where $S_{\hbar}$ is expanded out in powers of $X= (l_P/l_0)^2 = \frac{\hbar G}{l_0{}^2}$:
\begin{equation}
S_\hbar =\frac{l_0{}^2}{G} \Big(  S_{(0)} + X S_{(1)} + \frac{X^2}{2!} S_{(2)} + \ldots  \Big). \label{expandS}
\end{equation}
By writing the Ansatz in the above way, at lowest order, the conventional Hamilton-Jacobi equation is replaced by one with an inverted potential. With certain convexity and coercivity properties imposed on the nonlinear potential, the authors prove that a global, smooth `fundamental solution' to this equation exists, and that higher order quantum corrections to it can be calculated through the integration of linear transport equations derived from the Schr\"{o}dinger equation. When applied to the one-dimensional anharmonic oscillator problems, this method was found to produce results that agree with those given by the conventional Rayleigh/Schr\"{o}dinger perturbation theory for the energy eigenvalues. Remarkably, this method was found to yield wave functions that more accurately capture the more-rapid-than-gaussian decay known to hold for the exact solutions to these problems. In this present paper, the Wheeler-DeWitt equation for the Bianchi IX models in the absence of matter is solved semi-classically using the same Ansatz as above. However, for the reasons alluded to in the previous section, the overall method needs to be tailored to our Wheeler-DeWitt equation, as shown below. A more mathematical treatment of how the method of Ref. \cite{MMM} differs from the one used in this work is given in Ref. \cite{VM14}.

\subsection{The Transport Equations}

We plug in the modified semi-classical Ansatz for the ground state solution (\ref{Ansatz1})$-$(\ref{expandS}) into the Wheeler-DeWitt equation~(\ref{WdW}): 
\begin{equation}
\hbar \Box S_\hbar +  B \hbar \frac{\partial S_{\hbar}}{\partial\alpha} -  \nabla \! S_\hbar \! \cdot\! \nabla S_\hbar + \!\left(\frac{l_0}{l_P}\right)^4\!\!\!\hbar^2\ue^{4\alpha}V(\beta_{\pm})=0. \label{GS}
\end{equation}
Collecting all the terms of Eq.~(\ref{GS}) that have the same power of $X=\frac{\hbar G}{l_0{}^2}$, we obtain the Euclidean-signature Hamilton-Jacobi equation (for $k=0$) and the set of transport equations for the $k^{\mathrm{th}}$ quantum correction terms (for $k\geq1$): 
\begin{eqnarray}
-\nabla S_{(0)} \!\! \cdot \!\! \nabla S_{(0)} + \!\ue^{4\alpha}V(\beta_{\pm})=0. \quad (k=0)\label{eHJE1} \\
\begin{split}
k \Box S_{(k-1)} + Bk\frac{\partial S_{(k-1)}}{\partial\alpha} &\\ -\sum^k_{n=0}\binom{k}{n} \nabla S_{(n)} \!\! &\cdot \!\! \nabla S_{(k-n)}= 0. \quad(k\geq1). \label{GSk1}
\end{split}
\end{eqnarray}
Given a suitably chosen solution $S_{(0)}$ of Eq.~(\ref{eHJE1}), we can integrate Eq.~(\ref{GSk1}) along the flow of $S_{(0)}$ order by order in $X$ and calculate the quantum correction terms $S_{(k)}$ to any precision we require. The methods of \cite{MMM}, slightly modified, can be used to prove the existence of globally smooth solutions to these equations to all orders \cite{VM14}.

\subsection{Euclidean-signature Hamilton-Jacobi Equation}
The zeroth-order transport equation~(\ref{eHJE1}) is the Euclidean-signature Hamilton-Jacobi equation. To see this, consider the same ADM Hamiltonian formalism applied to a Euclidean-signature metric of the diagonalized Bianchi IX:
\begin{equation}
\mspace{50mu}\ud s^2 \bigg|_{\mathrm{Euc.}} = +N^2 \ud t^2 + \frac{l_0{}^2}{6\pi} \ue^{2\alpha}\big(\ue^{2\beta}\big)_{ij} \sigma^i \sigma^j.
\end{equation}
In this case, we end up with the same classical Hamiltonian constraint as before, but with the opposite sign in front of the ``kinetic part'' involving conjugate momenta:
\begin{equation}
H_{\mathrm{Euc.}}  = \frac{G \ue^{-3\alpha}}{24 \pi r^3} \Big[ -\mathbf{p}\cdot\mathbf{p}+\frac{l_0{}^4}{G^2} \ue^{4\alpha} V(\beta_\pm) \Big]=0.
\end{equation}
To write the Euclidean-signature Hamilton-Jacobi equation, we replace $p_q$ with $\frac{l_0{}^2}{G}\frac{\partial S_{(0)}}{\partial q}$:
\begin{equation}
\frac{G \ue^{-3\alpha}}{24 \pi r^3} \Big(\frac{l_0{}^2}{G}\Big)^2 \Big[ -\nabla S_{(0)} \!\! \cdot \!\! \nabla S_{(0)}  + \!\ue^{4\alpha}V(\beta_{\pm})\Big]=0. \label{eHJE}
\end{equation}
Hence, substituting the modified Ansatz into Lorentzian-signature Bianchi IX Wheeler-DeWitt equation results in (at the classical level) the Euclidean-signature (or `inverted potential') Hamilton-Jacobi equation. Eq.~(\ref{eHJE}) has been solved by others in different contexts \cite{Graham,MoncriefR}. Several different solutions to Eq.~(\ref{eHJE}) exist, such as the Hartle-Hawking `no boundary' solution \cite{BarberoR}. In this paper, we limit our scope to the fundamental solution found by Moncrief \& Ryan \cite{MoncriefR}, which is often referred to as the `wormhole' solution \cite{DEath}: 
\begin{equation}
S_{(0)} = \tfrac{1}{6} \ue^{2\alpha+2\beta_+} \left( \ue^{-6\beta_+}+2\h \right).\label{wormhole}
\end{equation}
The advantage of using the `wormhole' fundamental solution is that we can solve for the explicit $T$-dependence of the Misner variables. While this is not a prerequisite for our modified semi-classical method to work, it will serve to make the application clearer.

Note that we are always solving the Wheeler-DeWitt equation for the Lorentz-signature Bianchi IX models in Eq.~(\ref{WdW}). The fact that our zeroth-order transport equation in Eq.~(\ref{eHJE1}) (upon substituting our Ansatz as given in Eq.~(\ref{Ansatz1})) coincides with the Hamilton-Jacobi equation for the Euclidean-signature Bianchi IX models is merely an artifact of our method. It does not mean that we are at any point solving the Wheeler-DeWitt equation for the Euclidean-signature Bianchi IX models. In fact, the solution in Eq.~(\ref{wormhole}) would be a zeroth-order solution to the Euclidean-signature Bianchi IX models \emph{only if} it were exponentiated in the usual WKB way, $\ue^{iS/\hbar}$. But this is precisely where our method differs from what has been done in the past: we exponentiate Eq.~(\ref{wormhole}) according to our `modified Ansatz' $\ue^{-S/\hbar}$. Hence, Eq.~(\ref{wormhole}) is the zeroth-order solution to the Lorentz-signature Bianchi IX models, as we originally set out to do. 

\section{Results}
\subsection{Flow Equations: `Equations of Motion'}
From our Euclidean-signature Hamilton-Jacobi equation~(\ref{eHJE}), write the Hamilton equations for our Misner variables $q=\{\alpha, \beta_+,\beta_-\}$:
\begin{equation}
\frac{\ud q}{\ud t}=\frac{\partial (N H_{\mathrm{Euc.}})}{\partial p_q} \bigg|_{p_q = \frac{l_0{}^2}{G}\frac{\partial S_{(0)}}{\partial q}}.
\end{equation}
With a convenient time-gauge condition on the lapse function $N=-\sqrt{3}\,r\ue^{\alpha-2\beta_+}$, the `equations of motion' become:
\begin{eqnarray}
\dfrac{\ud \alpha}{\ud T}&= -\dfrac{1}{6}\bigg( \ue^{-6\beta_+}+2\h \bigg) \label{Alpha} \\
\dfrac{\ud \beta_+}{\ud T}&= -\dfrac{1}{3}\bigg( \ue^{-6\beta_+}-\:\: \h \bigg)  \\
\dfrac{\ud \beta_-}{\ud T}&=\;\; \dfrac{1}{\sqrt{3}}\sinh2\sqrt{3}\beta_-.\quad\qquad\qquad \label{Beta}
\end{eqnarray}
For convenience, we define the time-parameter $T=\sqrt{3}(t-t_0)$. Our choice of $t$ (and therefore $T$) is such that, while describing an expanding universe, $t$ runs from some finite $t_0>0$ in a backwards direction to $-\infty$. This is consistent with the negative sign in our choice of the lapse as $N=-\sqrt{3}\,r\ue^{\alpha-2\beta_+}$. A word of caution is in order. The `equations of motion' (\ref{Alpha})$-$(\ref{Beta}) are mere artifacts of our use of modified form of the Ansatz in Eq.~(\ref{Ansatz1}); they are not to be confused with equations of motion for our original Lorentz-signature Mixmaster models. The Euclidean-signature equations of motion (\ref{Alpha})$-$(\ref{Beta}) allow us to calculate higher order quantum corrections $S_{(k)}$ by defining the unique curve along which we can integrate the transport equations (\ref{GSk1}). In other words, we may write $\nabla S_{(0)}\!\cdot\!\nabla$ in Eq.~(\ref{GSk1}) as:
\begin{equation}
\nabla S_{(0)}\!\cdot\!\nabla  = 2\ue^{2\alpha+2\beta_+} \frac{\ud}{\ud T}. \label{flow}
\end{equation}
From here on, Eq.~(\ref{Alpha})$-$(\ref{Beta}) will thus be referred to as `flow equations.'

\subsection{Comparison with Belinskii, Gibbons et al.}\label{BGPP}
Having obtained our equations of motion from the Euclidean-signature Hamilton-Jacobi equation, we now compare with the results found in the 1978 paper titled ``Asymptotically Euclidean Bianchi IX metrics in Quantum Gravity'' by Belinskii et al \cite{GibbonsPage}. They write line element for the Euclidean-signature Bianchi IX models using $\omega_i$ variables:
\begin{equation}
\ud s^2 = (\omega_1\omega_2\omega_3) \ud \eta^2 +  \begin{pmatrix} \frac{\omega_2 \omega_3}{\omega_1} &0&0\\0&\frac{\omega_1 \omega_3 }{ \omega_2 }&0\\0&0&\frac{\omega_1 \omega_2}{ \omega_3}  \end{pmatrix}_{\!\!ij} \sigma^i \sigma^j.
\end{equation}
We can read off their variables $\omega_1, \omega_2,$ and $\omega_3$, which relate to our $\alpha$ and $\beta_\pm$ as,
\begin{equation}
\begin{split}
\omega_1 &= r^2 \ue^{2\alpha- \beta_+ - \sqrt{3}\beta_-} \\
\omega_2 &= r^2 \ue^{2\alpha- \beta_+ + \sqrt{3}\beta_-} \\
\omega_3 &= r^2 \ue^{2\alpha + 2\beta_+}.\qquad
\end{split}
\end{equation}
In contrast to our choice of lapse $N=-\sqrt{3}\,r\ue^{\alpha-2\beta_+}$, Belinskii et al. choose a different time lapse of $\tilde{N} =r^3 \ue^{3\alpha}$ (denoted by a different time-coordinate, $\eta$). Our equations of motion (\ref{Alpha})$-$(\ref{Beta}) are in full agreement with their equations of motion for $\omega_i$:
\begin{equation}
\frac{\ud \omega_1}{\ud \eta}  =  \omega_2 \omega_3 \qquad
\frac{\ud \omega_2}{\ud \eta}  =  \omega_3 \omega_1 \qquad
\frac{\ud \omega_3}{\ud \eta}  =  \omega_1 \omega_2.
\end{equation}
Solutions to these equations involve elliptic functions in $\eta$; similarly, we will see later that transport equations for $S_{(2)}$ and higher quantum correction terms all involve elliptic integrals. Our formalism and choice of lapse have the advantage that the fundamental solution and the first order quantum correction term can be expressed analytically in terms of elementary functions.

\subsection{First Order Transport Equation}

Putting $k=1$ in Eq.~(\ref{GSk1}), one can use the remarkable property $\Box S_{(0)} = 12S_{(0)}$ \cite{MoncriefR} to write the first order transport equation as:
\begin{equation}
12S_{(0)} + 2 B S_{(0)} - 2 \, \nabla S_{(0)} \!\! \cdot \!\!\nabla S_{(1)} = 0.
\end{equation}
Using Eq.~(\ref{flow}), we write the first order transport equation as an integral with respect to $T$:
\begin{equation}
\begin{split}
\ud S_{(1)} & = \frac{1}{2} \Big( B+6 \Big) \frac{S_{(0)}}{\ue^{2\alpha+2\beta_+}}\ud T \\
&=- \frac{1}{2} \Big( B+6 \Big) \frac{\ud \alpha}{\ud T} \ud T.
\end{split}
\end{equation}
In the second line, we use the identity in Eq.~(\ref{Alpha}). The above equation integrates to give $S_{(1)} = - \frac{1}{2}(B+6)\alpha$ plus any function that is invariant along the flow generated by $S_{(0)}$. Since we shall incorporate the influence of such `constants-of-the-motion' in our discussion of excited states, we exclude them from consideration as quantum corrections to the ground state. This first order quantum correction (as well as all higher order quantum corrections) vanishes for a particular ordering ($B=-6$), in agreement with \cite{MoncriefR}.

\subsection{Higher Order Transport Equations}

Using Eq.~(\ref{flow}), the transport equations for $k\geq2$ now read:
\begin{equation}
\begin{split}
-4 \ue^{2\alpha+2\beta_+}\frac{\ud S_{(k)}}{\ud T}  +k\Box S_{(k-1)} \\+ Bk\frac{\partial S_{(k-1)}}{\partial\alpha} -\sum^{k-1}_{n=1}\binom{k}{n}\nabla S_{(n)}\!\cdot\!\nabla S_{(k-n)}  = 0. \label{GSk2}
\end{split}
\end{equation}
By integrating the transport equations order by order in $X=\frac{\hbar G}{l_0{}^2}$, we can solve for all the quantum correction terms to the ground state solution. The second and higher quantum correction terms involve elliptic integrals, and their global smoothness to all orders is discussed in Ref. \cite{VM14}. The main emphasis of this paper is to show the spectrum of discrete excited states and their quantum corrections for the Bianchi IX models, to which we now turn.

\subsection{Explicit $T$-dependence}
Before we delve into calculating the excited states, it will be useful to simplify some of the subsequent analysis by first obtaining the explicit $T$-dependence of our Misner variables $\{\alpha(T),\beta_+(T),\beta_-(T)\}$ in our convenient gauge of $N = - \sqrt{3} r \ue^{\alpha - 2\beta_+}$. To do so, we integrate the flow equations (\ref{Alpha})$-$(\ref{Beta}) along $T$ from $0$~to~$-\infty$:
\begin{eqnarray}
\ue^{12\alpha}(T) & = & \ue^{12\alpha_0-6\beta_{+0}}H_+ \big( h_+ \, h_- \big)^2 \label{AlphaT} \\
\ue^{6\beta_+}(T) & = & \frac{H_+}{ h_+ \, h_-}\\
\ue^{2\sqrt{3}\beta_-}(T)&=&\frac{h_+}{h_-}.\label{BetaT}
\end{eqnarray}
In the above equations, the following abbreviations were used:
\begin{eqnarray}
\begin{split}
H_+ & = \ue^{6\beta_{+0}} - \cosh{2\sqrt{3}\beta_{-0}}+\tfrac{1}{2}(h_+^2+h_-^2) \\
& = \ue^{6\beta_{+0}} + (h_\pm)^2 - (h_{\pm0})^2 
\end{split}\\
h_{\pm} = \hpm. \mspace{10mu}
\end{eqnarray}
The initial values of the Misner variables at $t=t_0$ (or equivalently, at $T=0$) are denoted by $\{\alpha_0,\beta_{+0},\beta_{-0}\}$. We will also find the following identities useful in the next section:
\begin{eqnarray}
\h(T) & = & \frac{h_+^2+h_-^2}{2 h_+\,h_-}\\
\ue^{2\alpha+2\beta_+}& =& \ue^{2\alpha_0-\beta_{+0}}\sqrt{H_+}\\
\ue^{4\alpha-2\beta_+}& = & \ue^{4\alpha_0-2\beta_{+0}}h_+h_-.
\end{eqnarray}

\subsection{Excited States}

\begin{figure}[htb]
\begin{minipage}[b]{0.45\linewidth}
\centering
\includegraphics[width=\textwidth]{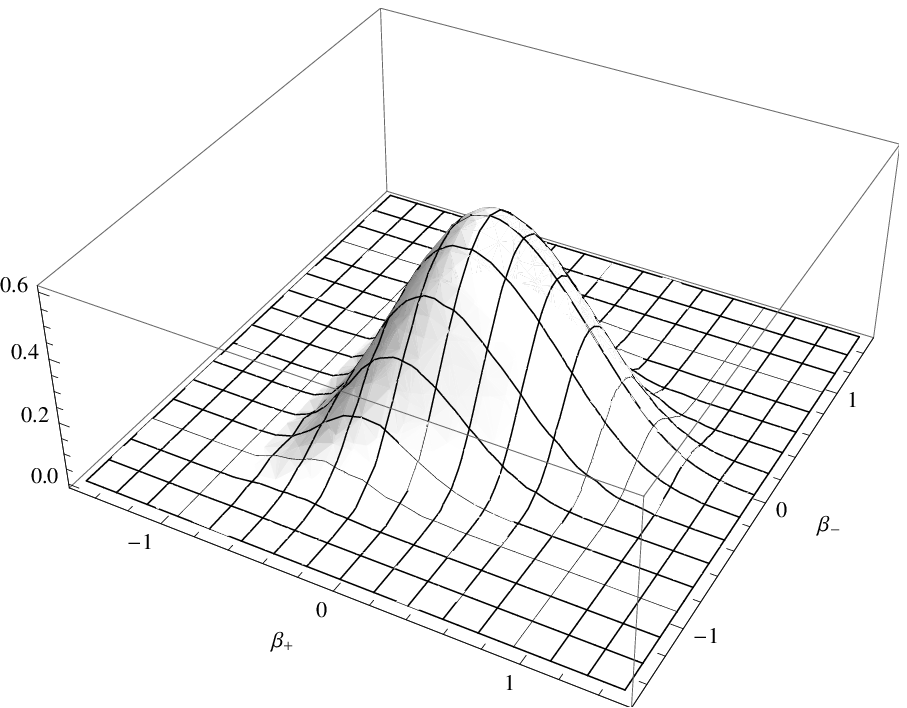} 
\caption{A plot of $\Psi_{(0,0)}$, with $\alpha=0$.}
\label{Psi00}
\end{minipage}
\hspace{0.5cm}
\begin{minipage}[b]{0.45\linewidth}
\centering
\includegraphics[width=\textwidth]{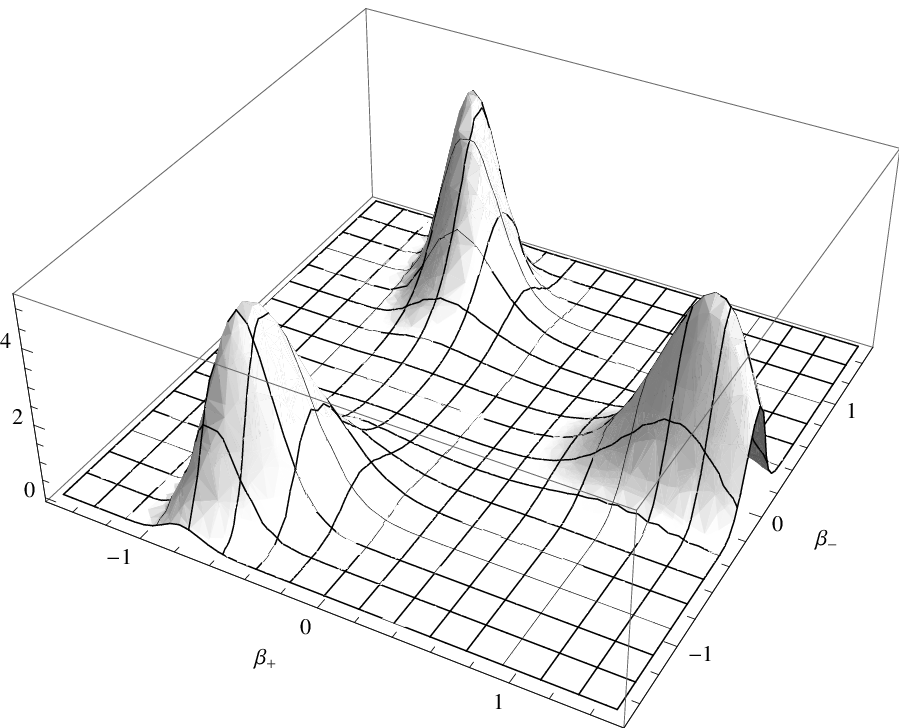} 
\caption{A plot of $\Psi_{(2,0)}=\Psi_{(0,2)}$, with $\alpha=0$.}
\label{Psi20}
\end{minipage}
\end{figure}

\begin{figure}[htb]
\begin{minipage}[b]{0.45\linewidth}
\centering
\includegraphics[width=\textwidth]{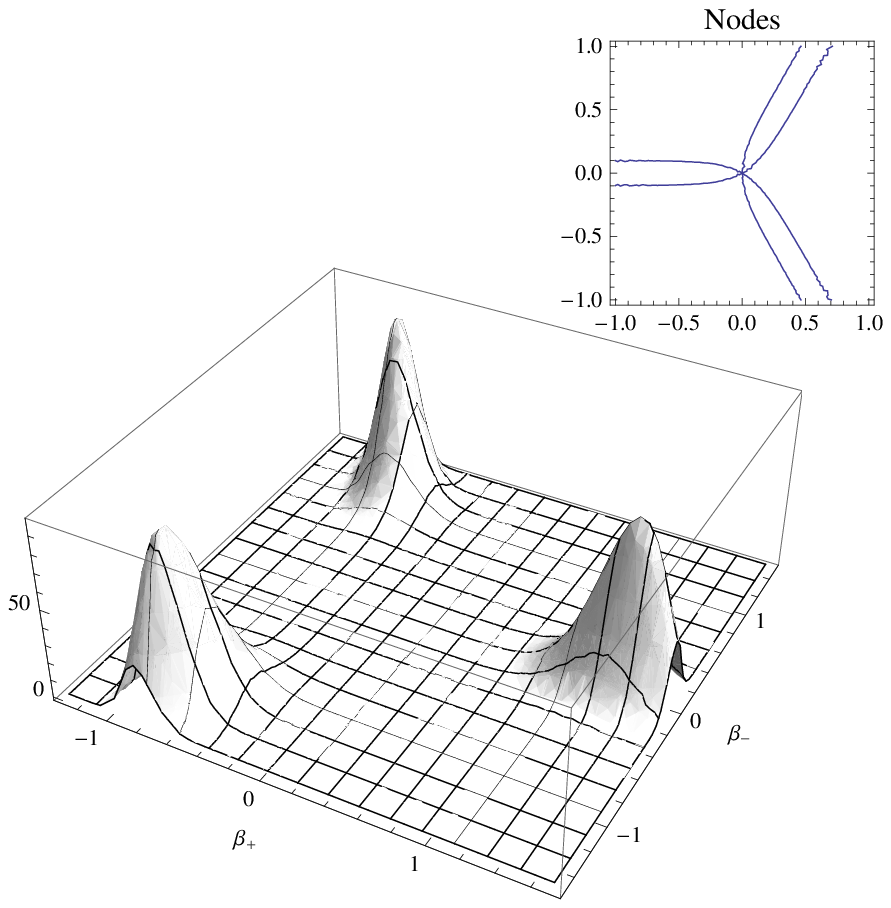} 
\caption{A plot of $\Psi_{(3,0)}=-\Psi_{(1,2)}$, with $\alpha=0$. Inset: Nodes of the wave function in $\beta$-space.}
\label{Psi30}
\end{minipage}
\hspace{0.5cm}
\begin{minipage}[b]{0.45\linewidth}
\centering
\includegraphics[width=\textwidth]{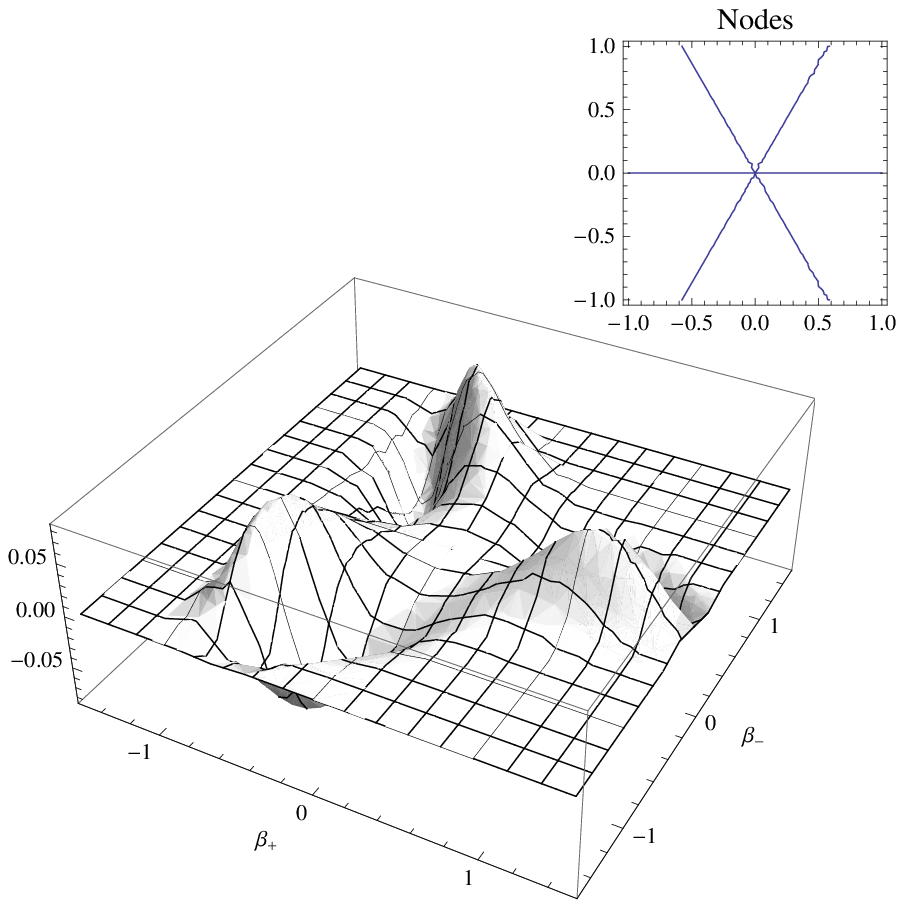} 
\caption{A plot of $\Psi_{(0,3)}=-\Psi_{(2,1)}$, with $\alpha=0$. Inset: Nodes of the wave function in $\beta$-space.}
\label{Psi03}
\end{minipage}
\end{figure}

For the excited states, we substitute the Ansatz:
\begin{equation}
\Psi = \phi_\hbar \ue^{-S_\hbar / \hbar}\label{exAnsatz}
\end{equation}
in the same Wheeler-DeWitt Eqution (\ref{WdW}), and we expand the wave function $\phi_\hbar$ in powers of $X=\frac{\hbar G}{l_0{}^2}$, as before:
\begin{equation}
\phi_{\hbar}=\phi_{(0)}+X \phi_{(1)} + \dfrac{X^2}{2!}\phi_{(2)} + \cdots + \dfrac{X^k}{k!}\phi_{(k)}+ \cdots .
\end{equation}
The function $S_\hbar$ is \emph{chosen} to coincide with what was previously defined for the ground state:
\begin{equation}
S_\hbar =\frac{l_0{}^2}{G} \Big(  S_{(0)} + X S_{(1)} + \frac{X^2}{2!} S_{(2)} + \ldots  \Big). 
\end{equation}
Note that for the excited states, the $\phi_\hbar$ function will have nodes where they become zero \cite{MMM}. As such, we cannot `absorb' $\phi_\hbar$ into $S_\hbar$ and re-express them as some quantum corrections to $S_{(0)}$; doing so would cause $S_{\hbar}$ to become infinite at those nodes and thus contradict our expansions of the Ansatz in powers of $\hbar$. By construction then, the $S_\hbar$ in the excited state solution satisfies Eq.~(\ref{GS}), and the Wheeler-DeWitt Equation now takes the form:
\begin{equation}
\hbar \Big(\Box\phi_{\hbar} + \! B \frac{\partial \phi_{\hbar}}{\partial\alpha} \Big)\! -2 \nabla S_{\hbar} \! \cdot \! \nabla \phi_\hbar =0 . \label{ES}
\end{equation}
At zeroth order, the equation for $\phi_{(0)}$ becomes:
\begin{equation}
\begin{split}
-2\frac{l_0{}^2}{G}\nabla S_{(0)} \! \cdot \! \nabla \phi_{(0)}& = 0 \\
 \frac{\ud\phi_{(0)}}{\ud T}& = 0, \label{phi0}
\end{split}
\end{equation}
where we use the identity in Eq.~(\ref{flow}) in the previous section. In other words, we want to find a combination of $\alpha$, $\beta_+$, and $\beta_-$ that gives a function invariant in $T$. From their explicit $T$-dependences, we define a pair of functions that satisfy $\frac{\ud\Phi_\pm}{\ud T}=0$:
\begin{equation}
\Phi_\pm=\ue^{4\alpha-2\beta_+}\big(\ue^{6\beta_+}-\ue^{\pm2\sqrt{3}\beta_-} \big).
\end{equation}
In order to maintain the natural $\frac{2\pi}{3}$ rotational symmetry in the overall solution, we further define $C_0, C_1, C_2$ and $S_0, S_1, S_2$, which are related to each other under discrete rotations of $\frac{2\pi}{3}$:
\begin{eqnarray}
C_0 &= \frac{1}{12} \Big(\Phi_+ + \Phi_- \Big) \quad & S_0 = \tfrac{1}{4\sqrt{3}} \Big(\Phi_- - \Phi_+ \Big) \\
C_1 &= \frac{1}{6} \Big(- \Phi_+ + \frac{1}{2} \Phi_- \Big) \quad& S_1 = -\tfrac{1}{4\sqrt{3}} \Phi_- \\
C_2 &= \frac{1}{6} \Big(- \Phi_- + \frac{1}{2} \Phi_+ \Big) \quad& S_2 = \tfrac{1}{4\sqrt{3}} \Phi_+.
\end{eqnarray}
Using the above functions, we may define a family of approximate solutions that satisfy Eq.~(\ref{phi0}) and are invariant under rotations of $\frac{2\pi}{3}$:\begin{equation}
\begin{split}
\Psi_{(m,n)}& = \phi^{(m,n)}_{(0)} \ue^{-\frac{l_0{}^2}{l_P{}^2}S_{(0)}}\\&= \tfrac{1}{3}\Big( C_0{}^mS_0{}^n + C_1{}^mS_1{}^n + C_2{}^mS_2{}^n \Big) \ue^{-\frac{l_0{}^2}{l_P{}^2}S_{(0)}}. \label{Phimn}
\end{split}
\end{equation}
Hence, the leading order excited state solutions are labeled by a pair of positive integers $(m,n)$ that can be plausibly interpreted as graviton excitation numbers for the two independent anisotropy degrees of freedom. To see why this interpretation seems natural, note first that, to leading order,
\begin{equation}
\ue^{-S_\hbar/\hbar} \sim \ue^{-\frac{l_0{}^2}{l_P{}^2}S_{(0)}} \sim \ue^{-\frac{l_0{}^2}{l_P{}^2}\ue^{2\alpha}(\frac{1}{2}+2(\beta_+{}^2 + \beta_-{}^2)+\cdots)},
\end{equation}
which thus behaves, at any fixed $\alpha$, like a Gaussian near the origin in $\beta$-space (more and more sharply peaked the larger $\alpha$ becomes). Next note that, near the origin in $\beta$-space,
\begin{equation}
C_0{}^mS_0{}^n \sim \ue^{4(m+n)\alpha}(\beta_+{}^m\beta_-{}^n+\cdots),\label{CSexpand}
\end{equation}
which, again for fixed $\alpha$, has the form of the top order term in the product of Hermite polynomials that one would expect to see in the case of actual harmonic oscillator wave functions \footnote{In reference \cite{MMM} it was shown that, for actual harmonic oscillators, the higher order excited state corrections precisely fill in the lower order terms in the corresponding Hermite polynomials and then terminate, reproducing the exact solutions.}. Thus, at least near the origin in $\beta$-space, our excited state wave functions have features corresponding to those for actual harmonic oscillators at each fixed $\alpha$. Figures~\ref{Psi00}$-$\ref{Psi03} show 3D Plots of $\Psi_{(0,0)}$, $\Psi_{(2,0)}$, $\Psi_{(3,0)}$, and $\Psi_{(0,3)}$ in $\beta$-space. As expected, $\Psi_{(0,0)}$ reproduces our ground state solution. Interestingly, $\Psi_{(1,0)}=\Psi_{(0,1)}=\Psi_{(1,1)}=0$, and so the first non-zero excited solution is $\Psi_{(2,0)}  = \Psi_{(0,2)}$, whose node is at the origin in $\beta$-space. The nodes of the higher excited states $\Psi_{(3,0)}$ and $\Psi_{(0,3)}$ are shown in Figures~\ref{Psi30}$-$\ref{Psi03}.
\vspace{-1cm}
The ground state solution has a peak at the origin, corresponding to the classically-forbidden isotropic FRW metric without matter. For the excited state solutions, however, the most probable states are located away from the origin (anisotropic metrics). 
We can write the transport equations for higher-order quantum correction terms of the excited solutions:
\begin{equation}
\begin{split}
-4 \ue^{2\alpha+2\beta_+}\frac{\ud \phi^{(m,n)}_{(k)}}{\ud T}  +k\Box\phi^{(m,n)}_{(k-1)} \mspace{120mu}\\\mspace{40mu}+ Bk\frac{\partial \phi^{(m,n)}_{(k-1)}}{\partial\alpha} -2\sum^{k}_{r=1}\binom{k}{r}\nabla  S_{(r)} \! \cdot \! \nabla  \phi^{(m,n)}_{(k-r)} = 0. \label{ESk2}
\end{split}
\end{equation}

\section{Discussion}
The Bianchi IX models have been studied for over 50 years and remain a topic of study as recent as 2014 \cite{Damour14}. So far, no one has been able to calculate the explicit forms of the discrete spectrum of excited states for the Bianchi IX Wheeler-DeWitt equation, first conjectured by Misner in 1972 \cite{Misner72}. Employing a modified form of the semi-classical Ansatz, we constructed both the ground and the excited state solutions to the canonically quantized Mixmaster models. For the Moncrief-Ryan solution to the corresponding Euclidean-signature Hamilton-Jacobi equation, we showed how the ground state quantum correction terms associated with the operator ordering ambiguities in the Wheeler-DeWitt equation can be computed by integrating a set of linear transport equations along the flow of the Hamilton-Jacobi solution. As mentioned above, the (microlocal) methods of \cite{MMM} can be modified to prove the global smoothness of these quantum corrections to all orders \cite{VM14}.

We calculated the explicit, leading order forms of a family of excited state solutions labeled by a pair of positive integers that can be naturally interpreted as graviton excitation numbers for the two independent anisotropy degrees of freedom. The smoothness of these excited state correction terms can also be handled by a modification of the arguments given in \cite{MMM}. In order to help in the physical interpretation of these excites states, the author has submitted to the same journal another paper in which a subset of the Bianchi IX family of models, namely the Taub family, is investigated using a perturbative approach \cite{Bae2014b}.

\begin{acknowledgements}
I am grateful to Professor Vincent Moncrief for valuable discussions at every stage of this work. I would like to thank Yale University for financial support.
\end{acknowledgements}

\bibliography{MyFirstPaper}{}

\begin{comment}

\end{comment}

\end{document}